\documentclass[prl,twocolumn,showpacs,superscriptaddress]{revtex4}

\usepackage[dvips]{graphicx}
\usepackage{amsmath}
\usepackage{amssymb}
\usepackage{color}

\begin{document}

\preprint{}

\title{Tricritical Phenomena at the Cerium $\gamma \rightarrow \alpha$ Transition}

\author{J.C. Lashley}
\affiliation{
   Los Alamos National Laboratory,
   Los Alamos, NM 87545}
\author{A.C. Lawson}
\affiliation{
   Los Alamos National Laboratory,
   Los Alamos, NM 87545}
\author{J.C. Cooley}
\affiliation{
   Los Alamos National Laboratory,
   Los Alamos, NM 87545}
\author{B. Mihaila}
\affiliation{
   Los Alamos National Laboratory,
   Los Alamos, NM 87545}
\author{C.P. Opeil}
\affiliation{
   Los Alamos National Laboratory,
   Los Alamos, NM 87545}
\author{L. Pham}
\affiliation{
   Los Alamos National Laboratory,
   Los Alamos, NM 87545}
\author{W.L. Hults}
\affiliation{
   Los Alamos National Laboratory,
   Los Alamos, NM 87545}
\author{J.L. Smith}
\affiliation{
   Los Alamos National Laboratory,
   Los Alamos, NM 87545}

\author {G.M.~Schmiedeshoff}
\affiliation{Department of Physics,
   Occidental College,
   Los Angeles, CA 90041}

\author {F. R.~Drymiotis}
\affiliation{Department of Physics,
   Clemson University,
   Clemson, SC 29634}

\author {G. Chapline}
\affiliation{Lawrence Livermore National Laboratory,
   Livermore, CA 94551}

\author{S. Basu}
\affiliation{Physics Department,
   Temple University,
   Philadelphia, PA 19122}

\author{P.S. Riseborough}
\affiliation{Physics Department,
   Temple University,
   Philadelphia, PA 19122}

\begin{abstract}
The $\gamma \rightarrow \alpha$ isostructural transition in the
Ce$_{0.9-x}$La$_x$Th$_{0.1}$ system is measured as a function of La
alloying using specific heat, magnetic susceptibility, resistivity,
thermal expansivity/striction measurements. A line of discontinuous
transitions, as indicated by the change in volume, decreases
exponentially from 118 K to close to zero with increasing La doping
and the transition changes from being first-order to continuous at a
critical concentration $0.10 \leq x_c \leq 0.14$. At the tricritical
point, the coefficient of the linear $T$ term in the specific heat
$\gamma$ and the magnetic susceptibility start to increase rapidly
near $x$ = 0.14 and gradually approaches large values at $x$=0.35
signifying that a heavy Fermi-liquid state evolves at large doping.
Near $x_c$, the Wilson ratio, $R_W$, has a value of 3.0, signifying
the presence of magnetic fluctuations. Also, the low-temperature
resistivity shows that the character of the low-temperature
Fermi-liquid is changing.
\end{abstract}


\maketitle

Bridgman's discovery of the first-order volume collapse between the
$\gamma$- and $\alpha$-phases of cerium in the 1930's has proven one
of the most fundamental problems in critical
phenomena~\cite{Bridgman_31,Bridgman_48}. Subsequent studies carried
out at room temperature showed that the transition between the
high-temperature $\gamma$-phase and the low-temperature
$\alpha$-phase is isostructural (fcc-to-fcc) and is manifested by a
large decrease in volume (as much as~17\%) at a pressure 0.7
GPa~\cite{Lawson_49,Koskenmaki_78}. The magnetic susceptibility and
specific heat of the $\gamma$-phase are quite large and similar to
those expected of a system with local moments, while in the
$\alpha$-phase the susceptibility is Pauli-paramagnetic and the
coefficient of the linear $T$ term in the specific heat is unusually
small (12.8 mJ K$^{-2}$mol$^{-1}$) for a compound with non-localized
4f electrons~\cite{Koskimaki_75}.

The puzzling properties of $\gamma \rightarrow \alpha$ phase
transition in elemental cerium has lent itself to many theoretical
models based on often contradictory interpretations of the
experimental data. First, L. Pauling~\cite{Pauling_quote} and W.~H.
Zachariasen~\cite{Zachariasen_quote} independently suggested that in
the phase transition the f electrons responsible for the magnetic
properties are squeezed into the valence band. In this picture, the
volume collapse occurred as a result of the increased bonding by the
valence electrons. This was thought to occur since the wavefunctions
of the bonding electrons have smaller radii. The Pauling-Zachariasen
promotional model was challenged when Gustafson \emph{et
al.}~\cite{Gustafson_69,Gustafson_72} performed positron lifetime
and angular correlation measurements of the annihilating photons in
$\gamma$- and $\alpha$-Ce that indicated that there was no
significant change in the number of f electrons. These observations
were supported by Compton scattering data~\cite{Lengeler_80b}, as
well as x-ray absorbtion measurements of the $L$
edges~\cite{Lengeler_83}, which also showed no substantial valence
fluctuation at the transition. In view of the experimental evidence,
Johansson~\cite{Johansson_74} concluded that the promotional model
was incorrect and suggested that the $\gamma \rightarrow \alpha$
phase transition was a Mott transition in which the localized
magnetic 4f electron states were transformed into a broad
non-magnetic 4f band of Bloch states. However, photoemission
experiments~\cite{Allen_81,Croft_81} showed that in both phases the
f level was primarily located at an energy between 2 and 3~eV below
the Fermi energy, contradictory to what Johansson envisaged where,
in the itinerant state, the 4f band should cut across the Fermi
energy. Recently, this conclusion has been reinforced by magnetic
form factor~\cite{Murani_05} and phonon densities of
states~\cite{Manley_03} measurements that showed that the magnetic
moments remain localized in both phases.

On the basis of photoemission measurements, Allen and Martin
suggested that the phase transition was due to a Kondo volume
collapse~\cite{Allen_82}. Here, the $4f$ level is always below the
Fermi energy and results in a localized $4f$ magnetic moment. The
transition is due to the competition between the entropy of the
six-fold degenerate magnetic ion in the high-temperature state and
the binding energy of a Kondo singlet ($\sim k_B T_K$) formed in the
low-temperature state by binding a conduction electron to the local
moment. One of the signatures of the Kondo model is that the Wilson
ratio, i.e. the dimensionless ratio of the susceptibility to the
specific heat, $R_W$
\begin{equation}
   R_W \ = \ \frac{4 \, \pi^2 \, k_B^2}{3 \, (g \mu_B)^2} \
             \frac{\chi(0)}{\gamma}
   \>,
\end{equation}
should have a value intermediate between $2$, for a spin one-half
model, and unity, as expected either in the limit of large spin
degeneracy $(2j+1) \gg 1$ or for non-interacting electrons.

\begin{figure}[t]
\includegraphics[width=\columnwidth]{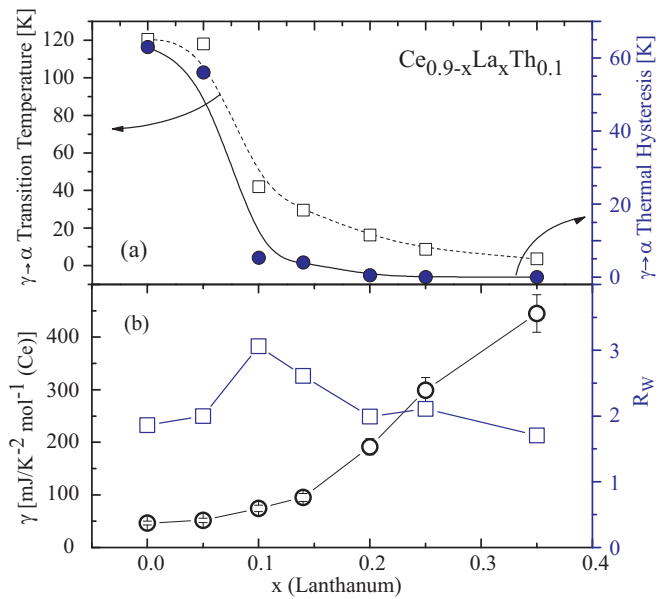}
\caption{(a) The effect of alloying on the $\gamma \rightarrow
\alpha$ transition is shown along the left y-axis, whereas the
thermal hysteresis is shown on the right y-axis. The transition
temperature decreases exponentially, and the hysteresis goes to zero
(within experimental error) at $0.10 \leq x_c \leq 0.14$, signifying
a tricritical point. (b) The variation in the electronic specific
heat $\gamma$ is shown on the left y-axis, whereas the $R_W$ is
shown on the right y-axis. Near $x_c$ one sees a peak in $R_W$,
signifying the onset of a heavy Fermi-liquid state.} \label{Figure1}
\end{figure}

Although in elemental Ce, the vibrational entropy change per atom is
estimated to account for about half the total entropy
change~\cite{Darling_04}, the entropy change due to phonon softening
at the transition of doped materials is small~\cite{Manley_02}.
Therefore, the entropy of the local moments of the high-temperature
phase must play an important role in driving the transition.
Recently, Dzero \emph{et al.} have suggested that, if the
low-temperature phase is non-magnetic and the high-temperature phase
has local moments, then application of an magnetic field, $B$,
should result in a depression of the critical
temperature~\cite{Dzero_00}. Dzero \emph{et al.} predicted that the
phase-boundary in the $(B,T)$-plane should have a semi-elliptical
form. Since doping with Th was known to suppress an unwanted portion
of the $\beta$-phase and also reduce the transition temperature, it
was possible to find a doping level at which the phase boundary
could be measured in the accessible field ranges. The observed phase
boundary~\cite{Lashley_05} was found to be consistent with the
predictions of Dzero \emph{et al.}~\cite{Dzero_00}, albeit with
small modifications due to the cubic crystal field splitting. For a
range of higher doping concentrations, it was shown that there was a
first-order transition line segment with a critical point at each
end~\cite{Smith_83,Smith_84,Lashley_05}. At a critical
concentration, the two critical points coalesce, and for higher
concentrations the transition is purely continuous.

In this paper we show that, with an optimal level of doping, it is
possible to push the transition temperature close to 0~$K$, so that
the $\gamma \rightarrow \alpha$ phase transition occurs as a
tricritical point. Furthermore, as the critical concentration is
approached, under ambient pressure, the coefficient $\gamma$ of the
linear $T$ term in the heat capacity shows a rapid increase, and is
still increasing on the La-rich side of the transition. The magnetic
susceptibility follows the same trend as the concentration is
varied. This suggests that the transition occurs from a state
characterized by pudgy electronic masses to a heavy fermion state.
However, the Wilson ratio peaks at a value above 2 for a narrow
range of concentrations where the specific heat and susceptibility
vary most rapidly with the doping concentration. That is, there is
an increase in the Wilson ratio at the critical concentration as
determined from the volume change.

\begin{figure}[b]
\includegraphics[width=\columnwidth]{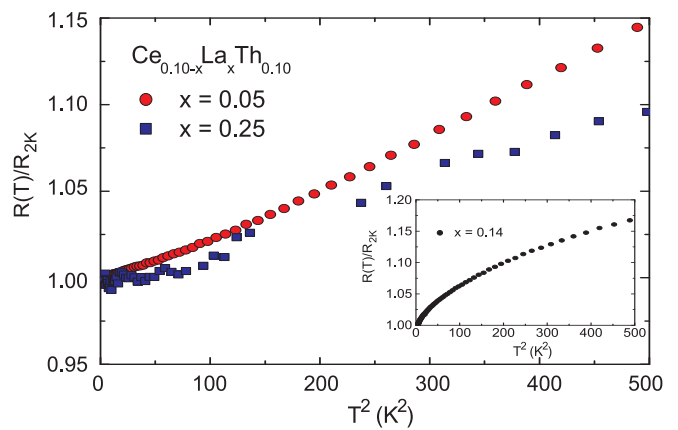}
\caption{Temperature variation of the resistivity for concentrations
above and below~$x_c$. For convenience, the resistivity is scaled
with respect to the resistivity measured at T=2K, $R_{2K}$, which
has the values of $3.4\times 10^{-4}\,\Omega$cm and $9.14\times
10^{-5}\,\Omega$cm for x=0.25 and x=0.05, respectively. The inset
shows $R(t)/R_{\mathrm{2K}}$ for x=0.14, near $x_c$.} \label{fig_T2}
\end{figure}

\begin{figure}[t]
\includegraphics[width=\columnwidth]{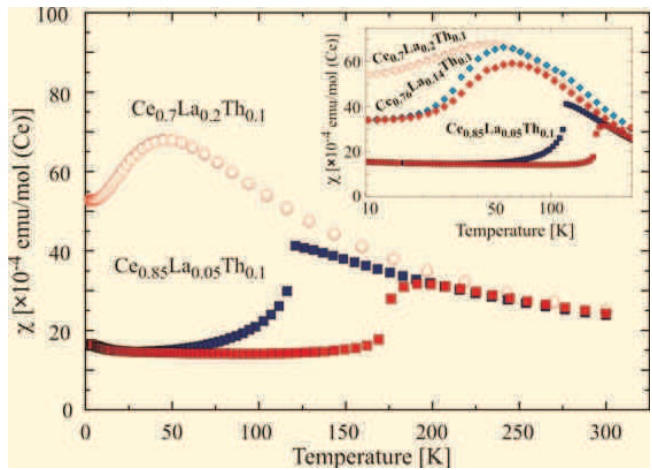}
\caption{The magnetic susceptibility as a function of temperature is
shown for La concentrations below and above $x_c$. We note the
discontinuous nature of the transition for $x$ = 0.05, whereas the
transition features a continuous cusp at $x$=0.20. Also, we note the
thermal hysteresis phenomenon present below $x_c$, disappears above
$x_c$. In order the emphasize the $\alpha$ phase, the transition
from Curie-Weiss behavior (high-temperature $\gamma$ phase) into a
Pauli temperature-independent susceptibility is shown on a
logarithmic scale in the inset. The inset also shows the
susceptibility for a third composition near~$x_c$.} \label{Figure2}
\end{figure}

Samples are prepared by arc melting the 99.99 $\%$ metals in an
atmosphere of Ar. The ingot is melted several times on each side and
then sealed in quartz and annealed for 5 days at 480 C. Chemical
composition of the ingots were made by inductively coupled mass
spectrometry. Specific heat, magnetic susceptibility, and thermal
expansivity were measured on a Quantum Design Physical Properties
Measurement System (PPMS). Specific heat was measured by a thermal
relaxation technique. Given the large volume change on cooling it
was necessary to fabricate a spring-loaded copper screw in a copper
sample holder. The holder was attached to the platform with a thin
layer of Apiezon N grease. The specific heat of the copper sample
holder and the specific heat of Apiezon N were measured separately
and subtracted from the total specific heat to obtain the specific
heat of the sample. Measurements were performed below 20 K. The
magnetic susceptibility was measured in a vibrating sample
magnetometer (VSM) in fields up to 9 T with a frequency of 40 Hz.
The coefficient of linear thermal expansion $\alpha$ was measured in
a three-terminal capacitive dilatometer (see Ref.~\cite{Lashley_06}
for details of the apparatus).

Figure~\ref{Figure1}(a) shows the suppression of the transition
temperature (as measured in the thermal expansion) and the thermal
hysteresis (as measured in the magnetic susceptibility) as a
function of La doping. One sees a crossover from a line of
discontinuous transitions, where the hysteresis is 5 K at $x \leq
0.14$ to a line of continuous transitions at $x \geq 0.14$. This
result is mirrored in the electronic specific heat and Wilson ratio
as shown in Fig.~\ref{Figure1}(b). Here, a low-temperature heavy
Fermi-liquid state develops for concentrations greater than $x_c$,
where the transition temperature of the continuous $\gamma
\rightarrow \alpha$ is close to zero. The Wilson ratio indicates
that the heavy Fermi-liquid state does not evolve directly from the
$\alpha$ Fermi-liquid phase, as it shows significant deviations from
the value of 2 for a small range of concentrations around the
critical concentration. This finding is supported by the
low-temperature variation of the resistivity which shows a
$T^2$-like dependence for concentrations above and below $x_c$ as is
expected for a Fermi liquid (see Fig.~\ref{fig_T2}). The
Kadowaki-Woods $x \leq x_c$ is in agreement with the expected value,
but a good $T^2$ fit could not be obtained for $x \geq x_c$.
However, for concentrations near $x_c$ the resistivity varies
linearly with $T$ down to 1.9 K which was the lowest temperature
measured. This type of change in the temperature variation of the
resistivity is frequently found near quantum critical points, which
is what is expected when a critical point is moved to zero.

The magnetic susceptibility data for compositions spanning $x_c$ are
shown in Fig.~\ref{Figure2}. These data were recorded at a magnetic
field of H = 9~T with a cooling/warming rate of 0.2 K/min. For the
$x$ = 0.05 sample one sees a discontinuous change (at 115 K on
cooling) from Curie-Weiss behavior in the $\gamma$-phase to a
temperature-independent susceptibility below the transition.
Conversely, for $x$ = 0.20 one sees a continuous change with
negligible thermal hysteresis $\Delta$T ($\Delta$T$\leq$ 0.5 K). In
the $\alpha$-Ce phase of this sample, the susceptibility does not
return to a temperature-independent susceptibility which we take as
a sign of sluggish kinetics. We have observed sluggish kinetics in
the thermal expansion and in previous transport measurements in
magnetic fields~\cite{Lashley_05}. In the Fig.~\ref{Figure2} inset,
the susceptibility is shown on a log-T scale to emphasize the
low-temperature phase. Here a third composition is added to show the
evolution from a first-order to a continuous transition.

\begin{figure}[b]
\includegraphics[width=\columnwidth]{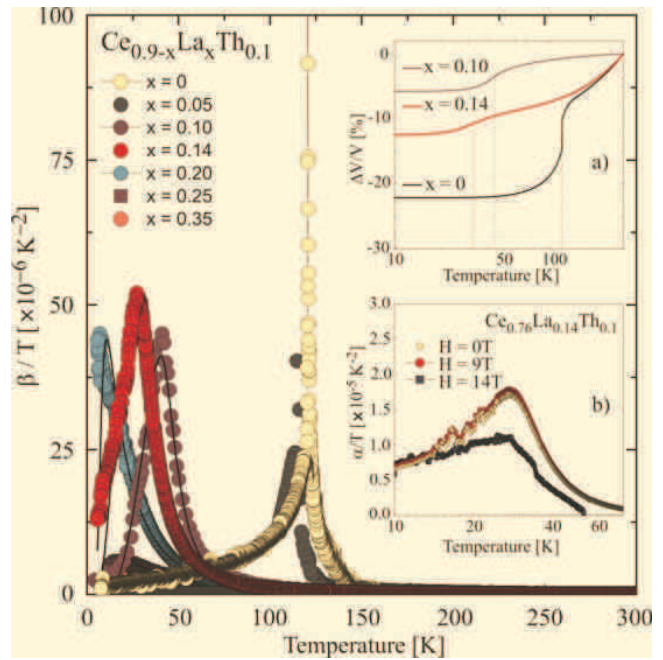}
\caption{The $\gamma \rightarrow \alpha$ transition as measured by
the thermal expansivity is shown as a function of $x$ in the main
figure. Mirroring the susceptibility behavior, the discontinuous
nature is apparent at compositions below~$x_c$. At compositions
above $x_c$ the volume expansivity, $\beta$, is fit to the A-P
model. Inset (a) shows also shows the evolution of the critical
point in $\Delta V/V$, while the magnetic field dependence is shown
in inset (b).} \label{Figure3}
\end{figure}

The phase boundary in the heavy Fermi-liquid phase was fit using the
Fermi-liquid model~\cite{Smith_83,Smith_84}, whereas in the
paramagnetic phase we used the two-level model proposed by Aptekar'
and Ponyatovskiy (A-P)~\cite{Aptekar_68a,Aptekar_68b} for the
isomorphous $\gamma \rightarrow \alpha$ transformation in cerium
metal. The crucial input to the Fermi-liquid model is the volume
dependence of the Fermi-liquid temperature. On defining the
Fermi-liquid temperature as being proportional to $\gamma^{-1}$, we
find that in the Fermi-liquid model the transition is driven by an
abrupt change in $\gamma^{-1}$ that occurs over a small range of
molar volumes. The cause for an abrupt variation of the Fermi-liquid
temperature might be considered outside the scope of the
Fermi-liquid model, but might be due to the existence of a critical
interatomic spacing for Ce~\cite{Hill_02}.

\begin{figure}[t]
\includegraphics[width=\columnwidth]{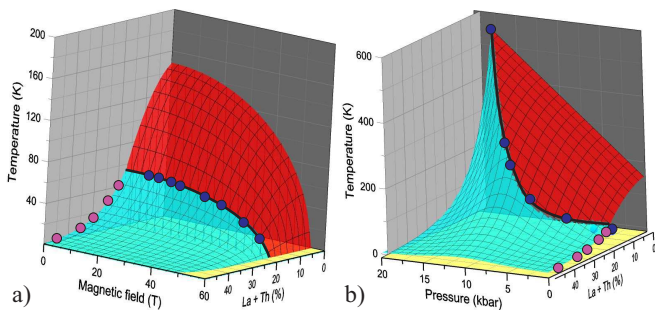}
\caption{a) Three dimensional phase diagram in temperature, magnetic
field, and composition space. The line of tricritical points
separating the low-temperature nonmagnetic states (red) and the
heavy Fermi-liquid states (blue), is described by the theory of
Dzero \emph{et al.}~\cite{Dzero_00}. b) Three dimensional phase
diagram in temperature, pressure, and composition space. The line of
tricritical points separating the low-temperature nonmagnetic states
(red) and the heavy Fermi-liquid state (blue) is obtained from the
pressure data points taken from Refs.~\cite{Smith_83,Smith_84},
combined with the zero-pressure points reported here to complete the
phase diagram.} \label{Figure4}
\end{figure}

Whereas the Fermi-liquid model describes the condensation of
electrons into a homogeneous state at low temperatures, the A-P
model is a thermodynamic model based on the notion of the
equilibrium in an inhomogeneous binary phase. Because the phase
boundary is determined by the free energy, it is possible in
principle to determine the energy difference between the two phases,
$\Delta$E, entropy difference $\Delta$S, volume difference $\Delta$V
and energy of mixing $U$, from knowledge of the temperature
dependence of the boundary. A practical procedure is to integrate
the experimental boundary between 10 and 300K to get $\Delta$V, and
find the remaining three parameters from the data using the
Levenberg-Marquardt optimization. A fit of the A-P model to the
measured thermal expansion is shown in Fig.~\ref{Figure3}. In this
figure the volume expansivity, $\beta$/T versus T is plotted to show
the suppression of the discontinuous $\gamma \rightarrow \alpha$
transitions. At La compositions $x\leq x_c$ one sees the first-order
character of the transition. Here the transitions are too sharp to
be fit by the A-P model. For $x\geq x_c$, one sees a continuous
transition. The evolution from first order to continuous is mirrored
in the volume shown in the inset (a) of Fig.~\ref{Figure3}. Here one
clearly sees the critical point at $x = 0.14$ shown by the gentle
cusp near 20 K. At this composition magnetic fields move the phase
boundary of the transition in $H^2$ $T^2$ space as can be inferred
from inset (b). Combining results from~\cite{Lashley_05} and the
current thermodynamic data, one finds the phase $T$-$H$-$x$ diagram
shown in Fig.~\ref{Figure4}a. Similarly, combining pressure results
from ~\cite{Smith_83} one finds a phase $T$-$p$-$x$ diagram shown in
Fig.~\ref{Figure4}b. In each case the surface marks the phase
boundary between the high-temperature local-moment phase and the
low-temperature Fermi-liquid phase. The tricritical line indicates a
change in the transition from first to second order. The tricritical
line occurs at concentrations where the character of the
low-temperature Fermi-liquid is rapidly changing, and the
resistivity shows changes characteristic to a system in the vicinity
of a quantum critical point.

In conclusion, in this paper we have shown that it is possible to
move the transition temperature close to 0~$K$, so that the $\gamma
\rightarrow \alpha$ phase transition occurs as a tricritical point.
At ambient pressure, the coefficient $\gamma$ of the linear $T$ term
in the heat capacity and the magnetic susceptibility show a rapid
increase as the critical concentration is approached. This increase
continues on the La-rich side of the transition. Based on these
observations, we suggest the transition occurs between a state
characterized by pudgy electronic masses and a heavy fermion state.
The Wilson ratio reaches a value above 2 for a narrow range of
concentrations near $x_c$, signifying the presence of magnetic
fluctuations. Here the character of the low-temperature Fermi-liquid
is rapidly changing, and the resistivity shows changes
characteristic to a system in the vicinity of a quantum critical
point.

%
%
\bibliography{cerium}

\end{document}